# Magnetic properties and magnetocaloric effects in NaZn$_{13}$-type La(Fe, Al)$_{13}$-based compounds*

Shen Bao-Gen, Hu Feng-Xia, Dong Qiao-Yan, Sun Ji-Rong

*State Key Laboratory for Magnetism, Institute of Physics, Chinese Academy of Science, Beijing 100190, China*

**Abstract**

In this article, our recent progresses about the effects of atomic substitution, magnetic field and temperature on the magnetic and magnetocaloric properties of the LaFe$_{13-x}$Al$_x$ compounds are reviewed. With the increase of aluminum content, the compounds exhibit successively an antiferromagnetic (AFM), a ferromagnetic (FM), and a mictomagnetic state. Furthermore, the AFM coupling of LaFe$_{13-x}$Al$_x$ can be converted to a FM coupling by substituting Si for Al, Co for Fe and magnetic rare-earth R for La, or introducing interstitial C or H atom. However, low doping levels lead to FM clusters embedded in AFM matrix, and the resultant compounds can undergo, under appropriate applied fields, first an AFM-FM and then a FM-AFM phase transition while heated, with significant magnetic relaxation in the vicinity of the transition temperature. The Curie temperature of LaFe$_{13-x}$Al$_x$ can be shifted to room temperature by choosing appropriate contents of Co, C or H, and strong magnetocaloric effect can be obtained around the transition temperature. For example, for the LaFe$_{11.5}$Al$_{1.5}$C$_{0.2}$H$_{1.0}$ compound, the maximal entropy change reaches 13.8 J/kg K for a field change of 0-5 T, occurring around the room temperature. It is 42% higher than that of Gd and, therefore, this compound is a promising room-temperature magnetic refrigerant.

*Project supported by the National Natural Science Foundation of China, the Key Research Program of the Chinese Academy of Sciences, the National Basic Research Program of China, and the Hi-Tech Research and Development program of China.

# 1. Introduction

Refrigeration plays a very important role in various fields such as industrial and agricultural production, scientific research, aerospace, medicine, and daily life, and modern society is increasingly relying on refrigeration technology. The drawback of the conventional gas compression-expansion technology, which has been widely used today, is the high energy consumption and its adverse effect on environment. Therefore, exploring a new type refrigeration technology that is environment-friendly and energy-efficient becomes an urgent problem. Magnetic refrigeration based on the magnetocaloric effect (MCE) of magnetic materials has attracted worldwide attention because of its important application prospect.

The history for the research of magnetic refrigeration can be traced back to the year of 1881, when the MCE was first discovered by Warburg.[1] With the development of magnetic refrigeration in recent decades, magnetic refrigeration has been widely applied in the low temperature range and, meanwhile, the research on magnetic refrigeration has been gradually promoted from the low temperature range to high temperature range. In 1976, Brown observed a large MCE in rare earth Gd (its Curie temperature $T_C$ is 293 K) near the room temperature.[2] Since then Gd was ever thought to be the sole refrigerant for the magnetic refrigerators working around room temperature. The maximal magnetic entropy changes ($\Delta S$) for Gd are only 5.0 and 9.7 J/kg K for the field changes of 0-2 T and 0-5 T, respectively.[3] Furthermore, the phase transition temperature cannot be adjusted on demand. Many scientists over the world have devoted their attention to the exploration of new MCE materials with better performance than that of Gd, and important progress has been obtained in recent decades. In 1997, for example, Pecharsky et al. from the Ames lab of the USA reported that the peak value of $\Delta S$ of the rare-earth intermetallic compound $Gd_5Si_2Ge_2$ ($T_C$ =274 K) reaches 18 J/kg K for a field change of 0-5 T.[4] In the same year, Guo et al. from the Nanjing University of China reported an entropy change of 5.5 J/kg K for a field change of 0-1.5 T for the manganite $La_{0.8}Ca_{0.2}MnO_3$ ($T_C$ =274 K).[5] In 2000, Hu et al. from Institute of Physics of the Chinese Academy of Sciences found that the Heusler alloy $Ni_{51.5}Mn_{22.7}Ga_{25.8}$ has a large $\Delta S$ of 4.1 J/kg K for a field change of 0-0.9 T, associated with the martensitic-austenitic phase transition at 197 K.[6] They also found that single crystal $Ni_{52.6}Mn_{23.1}Ga_{24.3}$ exhibits a $\Delta S$ of 18 J/kg K for a field change of 0-5 T at the transition temperature of 300 K.[7] At the end of 2001 and the beginning of 2002, the Kyoto University of Japan and the Amsterdam University of Netherlands reported, respectively, the $MnAs_{1-x}Sb_x$[8] and $MnFeP_xAs_{1-x}$[9] compounds that show great magnetic entropy change in the room temperature range.

$LaFe_{13}$ binary compound with the cubic $NaZn_{13}$-type structure does not exist. It is necessary to introduce Al or Si atoms in order to obtain stable $LaFe_{13}$-based compound. In the $NaZn_{13}$-type structure, La atoms occupy the 8a crystallographic site, and Fe atoms occupy the 8b and 96i sites. Fe atoms at the two sites are denoted as $Fe_I$ and $Fe_{II}$, respectively. Al or Si atoms substitute $Fe_{II}$ randomly. La atoms and the Fe atoms at $Fe_I$ site form a CsCl structure. The $Fe_I$ atoms are surrounded by an icosahedron formed by 12 $Fe_{II}$ atoms and the $Fe_{II}$ atoms are surrounded by other 9 $Fe_{II}$ atoms and 1 $Fe_I$ atom at the nearest neighbor sites.[10] Based on the analysis of the phase formation rule for the $NaZn_{13}$-type compounds, we successfully synthesized the $LaFe_{13-x}Si_x$ compounds with low Si content, getting a new category of magnetic materials with large MCE.[11,12] The structure, phase transition, magnetic properties and MCE of the $LaFe_{13-x}Si_x$ compounds have been systematically investigated by neutron diffraction, Mössbauer spectroscopy measurements, theoretical analyses based on first-principle calculations and etc. We found that large MCE is resulted from a negative thermal lattice expansion and field-induced itinerant-electron metamagnetic transition.[12-17] Substituting Co for Fe or introducing interstitial atoms of C/H can effectively affect the exchange interaction between transition elements. This makes the phase transition temperature of the compounds adjustable in a wide temperature range covering the room temperature in the meantime maintaining large MCE. The maximal entropy change of the typical $LaFe_{13}$-based compounds exceeds 12 J/kg K and 20 J/kg K for the field changes of 0-2 T and 0-5 T, respectively, occurring around room temperature.[17-25]

$LaFe_{13-x}Al_x$ compounds with $1.04<x<7.02$ can be stabilized in the $NaZn_{13}$-type crystalline structure.[10,26,27] Diverse magnetic ground states are observed in this system with the variation of Al content, including (i) a mictomagnetic state in the content range of $4.9 \leq x \leq 7.0$, (ii) a soft FM state for $1.8<x\leq 4.9$, and (iii) an AFM state for $1.0 \leq x \leq 1.8$. Due to the large Fe-Fe coordination numbers and short Fe-Fe atom distance, the AFM ordering is weak and can be easily driven into the FM state by external magnetic field, accompanying significant lattice expansion.[27,28] Further investigations revealed that other ways can also drive the AFM-FM switching, such as adjusting the ratio between Fe and Al, substituting Fe with Co or La with Ce, Pr or Nd, or introducing interstitial atoms of C or H.[29-40] When x locates at the critical content range between the AFM and FM states, a thermal driven FM-AFM phase transition occurs. Around this transition, large MCEs that satisfy the requirements of Ericsson-cycle refrigerator were observed.[33,41] The magnetic and magnetocaloric properties of $LaFe_{11.4}Al_{1.6-x}Si_x$, $La_{1-x}R_xFe_{13-y}Al_y$ (R=Ce, Pr, Nd), $LaFe_{13-x}Al_xZ_y$ (Z=H, C), and $LaFe_{13-x-y}T_xAl_y$ (T=Mn, Co) were also systematically investigated. In this paper, we will give a review on our recent progresses, emphasising on the effects of Al, La, and Fe site substitution and interstitial atomic effect, the effects of magnetic field and temperature on phase transition, magnetic and magnetocaloric properties of the compounds.

## 2. Influences of the Mn and Co substitution for Fe on the magnetic properties and magntocaloric effects of $LaFe_{13-x}Al_x$ compounds

In 2000, Hu et al.[10,30] firstly reported on the magnetic properties and magnetocaloric effects of $LaFe_{13-x}Al_x$-based compounds, and found that a small amount of introduction of Co can convert the AFM coupling to the FM one for the $LaFe_{13-x}Al_x$ compound with an AFM ground state. For example, $LaFe_{11.12}Co_{0.71}Al_{1.17}$ compound exhibits the typical FM behavior with a second-order phase transition at $T_C$=279 K, and large $\Delta S$ was found around $T_C$. The peak values of $\Delta S$ are about 4.6 and 9.1 J/kg K[10] under the field changes of 0-2 T and 0-5 T, respectively, which are comparable with the results of Gd (5.0 and 9.7 J/kg K, $T_C$=293 K[3]). For $La(Fe_{1-x}Co_x)_{11.7}Al_{1.3}$ compounds, $T_C$ exhibits a monotonic growth with increasing Co content. It locates at 197 K when $x$=0.02 and increases to 309 K when $x$=0.08. Figure 1 displays the temperature dependence of magnetic entropy change of $La(Fe_{1-x}Co_x)_{11.7}Al_{1.3}$ Although the maximum value of magnetic entropy change reduces slightly with the increase of Co content, it still retains considerable magnitude at room temperature. Furthermore, the proper Co to Al ratio can also drive the Curie temperature to room temperature while remaining large MCE. For instance, the $T_C$ of $LaFe_{10.88}Co_{0.95}Al_{1.17}$ is 303 K, and the peak values of $\Delta S$ are 4.5 and 9.0 J/kg K under the field changes of 0-2 T and 0-5 T, respectively. The large $\Delta S$ is originated from the high saturation magnetization and the sharp magnetization change around $T_C$. The average magnetic moment per magnetic atom reaches to 2.1 $\mu_B$ in $LaFe_{10.88}Co_{0.95}Al_{1.17}$ compound.[31]

The crystal structure and magnetic properties of the LaCo$_{0.7}$Fe$_{12.3-x}$Al$_x$ (1.5≤$x$≤4.68) compounds were investigated by Wang et al.[42] FM behavior was observed when the content of Al is increased from 1.5 to 4.68. With the increase of the Al content, lattice parameter grows monotonously whereas $T_C$ firstly increases and then, after a maximal value (315 K) at $x$=2.6, decreases. There are five different Fe-Fe distances between the nearest neighboring Fe atoms for NaZn$_{13}$-type structure: one between Fe$_I$ and Fe$_{II}$ and the rest four between Fe$_{II}$ and Fe$_{II}$. The variation of five Fe-Fe distances and the nearest neighboring number of Fe atoms with the change of Al content determines the variation trend of $T_C$.

Wang et al.[43] investigated the substitution effect of Mn for Fe on the magnetic properties of La(Fe$_{1-y}$Mn$_y$)$_{11.4}$Al$_{1.6}$, and found that the compounds keep the NaZn$_{13}$-type structure in the content range of 0≤$y$≤0.25 though lattice constant linearly increases with the content of Mn due to the large atomic radius of Mn compared with Fe. Meanwhile, the magnetic properties of LaFe$_{13-x}$Al$_x$ show a strong dependence on local coordination environment and lattice volume. For samples $y$=0.05, 0.10, a field-induced metamagnetic transition takes place, leading to a ferromagnetic order. However, when the Mn content exceeds $y$=0.15, long-range magnetic order disappears. As mentioned above, the substitution of Fe by Mn slightly enlarges the lattice and therefore the Fe-Fe atomic distance. As a result, the weak AFM coupling between Fe atoms changes to FM coupling. Meanwhile, the AFM coupling between Fe-Mn strengthens with increasing Mn content. The competition between the FM and AFM coupling leads to a spin-glass behavior. The spin-glass behavior prevails in the content range of 0.05≤$y$≤0.25, and increasing Mn content causes a reduction in freezing temperature.

## 3. Substitution effect of Si for Al on magnetic properties and magnetocaloric effect for the LaFe$_{13-x}$Al$_x$ compounds

Dong et al[44] studied the effect of substitution of Al by Si on the magnetic properties and the MCE of LaFe$_{11.4}$Al$_{1.6}$. It was found that the lattice parameter of LaFe$_{11.4}$Al$_{1.6-x}$Si$_x$ nearly linearly decreases with increasing Si content. Figure 2(a) presents magnetic phase diagram of LaFe$_{11.4}$Al$_{1.6-x}$Si$_x$ (0≤$x$≤1.6). The compounds are AFM when the Si content $x$≤0.20. With increasing temperature an AFM-paramagnetic (PM) transition occurs at the Neel temperature ($T_N$) that decreases with the increase of Si content. However, when Si content is x=0.22, a cusp appears in thermal magnetization curves below $T_N$ ~191 K (Fig. 2(b)). This is a result of regular distribution of FM clusters in AFM background. Further increasing Si content to 25≤x≤0.40 leads to a FM ordering at low temperatures. A FM-AFM and an AFM-PM transition appear sucessively upon warming. The FM-AFM transition temperature $T_0$ rapidly increases with Si content. As Si content 0.50≤$x$≤1.60, FM properties prevail, and only a FM-PM transition occurs at $T_C$ that increases with Si content. However, the change of saturation magnetization is rather small with the substition of Al by Si, indicating that the Fe moment is nearly unchanged.

Figure 2(b) displays the temperature-dependent resistivity of LaFe$_{11.4}$Al$_{1.6-x}$Si$_x$ (x=0.22). With the increase of temperature, the resistivity firstly increases, then suddenly drops at 75 K, and then linearly increases again until an abrupt jumps that is followed by a monotonic decrease. As external magnetic field is applied, the resistivity-temperature dependence is very similar to the case of $H$=0. With the increase of magnetic field, the critical temperature for the resistivity drop shifts to low temperatures while the one for the resistivity jump shifts to high temperatures. This fact manifests the notable effect of magnetic field on the FM clusters and transition temperatures. Compared with the magnetization-temperature relation, a correspondence between the two abrupt resistivity changes and the AFM-FM and FM-PM transitions can be found: The resistivity is the biggest in the AFM state, the lowest in the FM state, and intermediate in the PM state. These phenomena can be understood based on two-fluid model.[27]

According to the phase transition characteristics, Dong et al[44] chose representative compounds $LaFe_{11.4}Al_{1.6-x}Si_x$ (x=0.3, 0.8) and investigated their MCE. As seen in Fig.3, for $LaFe_{11.4}Al_{1.6-x}Si_x$ (x=0.3), two entropy change peaks corresponding to the AFM-FM and FM-PM transitions appear at $T_0$ and $T_N$, respectively. The peak value at $T_0$ reaches 25-30 J/kg K, but this is a false signal caused by Maxwell relation.[45,46] Under a relative high magnetic field, the two peaks merge into one table-like peak with a maximal value of 7.8 J/kg K and a temperature span of ~70 K. For $LaFe_{11.4}Al_{1.6-x}Si_x$ (x=0.8), entropy change peaks at 182 K, corresponding to the FM-PM magnetic transition. The maximal entropy change is 9.1 and 14.0 J/kg K for a field change of 0-2 T, 0-5 T, respectively. The large entropy change is caused by the specific composition, which locates at a critical point between first-order and second-order phase transitions.

Shen et al[47] investigated magnetic and magnetocaloric properties for $LaFe_{11.03}Co_{0.8}Al_{1.17-x}Si_x$ (0.47<x≤1.17), and found that only a FM-PM transition occurs for all compositions and $T_C$ decreases but $\Delta S$ increases with increasing Si content. Figure 4 presents ΔS as a function of temperature under different fields for $LaFe_{11.03}Co_{0.8}Al_{1.17-x}Si_x$. For the field changes of 0-2 T and 0-5 T, the maximal values of $\Delta S$ are, respectively, 4.5 and 8.8 J/kg K for x=0.47, and 11.0 and 18.5 J/kg K for x=1.17. The growth of $\Delta S$ is caused by the change of transition nature from second-order to first-order, and the concurrent enlargement of lattice expansion during magnetic transition, as well as the enhancement of itinerant electron metamagnetic behavior.[11,16]

## 4. Substitution Effect of magnetic rare earth *R* for La on the magnetic and magnetocaloric properties of LaFe$_{13-x}$Al$_x$

For the LaFe$_{13-x}$Al$_x$ intermetallics with a AFM ground state, the replacement of La by magnetic rare earth R will produce an obvious effect on crystal structure, magnetic and magnetocaloric properties. The work of Wang et al[40,48] indicated that the substitution ratio of La by light rare earth R generally cannot exceed 30% to keep a stable NaZn$_{13}$-type structure. As heavy rare earth R is introduced to replace La, single phase can be hardly obtained. This fact indicates that the R elements with a large radius favor the stabile NaZn$_{13}$-structure. Due to the lanthanide contraction, the replacement of La by magnetic rare earth R causes a lattice contraction, leading to a variation in the Fe-Fe, Fe-Al-Fe bond lengths and Fe-Al-Fe bond angles. Meanwhile *R*-Fe and *R*-*R* interactions are introduced. The combined effect results in the weakness of AFM behavior. As experimentally shown, increasing R content causes a slow low temperature shift of $T_N$ and a reduction of the critical field driving AFM-FM transition, the latter indicates that the field-induced AFM-FM transition becomes easier as La is replaced by magnetic R. The substitution effect on magnetic properties enhances along the sequence of Ce, Pr, and Nd, and Nd doping yields the strongest effect. Figure 5 displays magnetization curve measured at 5K for La$_{1-x}$R$_x$Fe$_{11.5}$Al$_{1.5}$ (*R*=Ce、Pr、Nd). The critical field to drive the AFM-FM transition is 4.1 or 3.6 T for the Ce or Pr substitution, and it is nearly zero for Nd case.

Dong et al[49] and Chen et al[39] reported the effect of the replacement of La by Pr on magnetic properties and phase transitions for LaFe$_{11.4}$Al$_{1.6}$. From thermal magnetization curves measured under a low field of $H$=0.01 T, they found that the $T_N$ locates at ~188 K and a cusp appears below $T_N$, similar to the case of LaFe$_{11.4}$Al$_{1.6-x}$Si$_x$ ($x$=0.22). This is caused by the existence of FM clusters in the AFM background. The only difference is the irregular distribution of FM clusters in the AFM background for La$_{0.8}$Pr$_{0.2}$Fe$_{11.4}$Al$_{1.6}$. From temperature-dependent magnetization (*M-T* curve) measured under different magnetic fields, one can learn that magnetic field-driven AFM-FM, FM-AFM transitions take place. The peak at low temperature in the *M-T* curve becomes table-like and its height grows, indicating the increase of the FM component with increasing magnetic field. Figure 6 shows the *M-T* curve measured at $H$=0.45 T for La$_{0.8}$Pr$_{0.2}$Fe$_{11.4}$Al$_{1.6}$, where ZFC, FCC, and FCW refer to the modes of zero-field cooling and measuring on heating, field cooling and synchronously measuring, and field cooling and measuring on heating, respectively. The FCC and FCW curves show obvious discrepancy around the FM-AFM transition, a typical feature of first-order phase transition. Moreover, all the *M-T* curves obtained via the ZFC, FCC, and FCW modes displays many discontinuous jumps and steps, manifesting the metastable characteristics of the magnetic state.

The magnetic metastable behavior is further studied through measuring time-dependent magnetization at different temperatures. Figure 7 presents the reduced magnetization of La$_{0.8}$Pr$_{0.2}$Fe$_{11.4}$Al$_{1.6}$ as a function of time, measured under a field of 0.45 T in the ZFC mode. At temperatures below the AFM-FM transition, the magnetization keeps nearly unchanged with time, i.e., the system is frozen in the AFM state. With the increase of temperature, relaxation behavior is obviously enhanced, and the magnetization increases significantly with time. This fact reflects the formation and growth of ferromagnetic phases. At temperatures near AFM-FM transition point, the magnetization shows discontinuous jumps and steps, a signature of the strong competition of the AFM and FM ordering. In the temperature range above the AFM-FM transition, however, the relaxation phenomena cannot be observed. It indicates that the AFM-FM transition is completed. For either the ZFC or the FCC mode, the magnetization-time dependence can be classified into two categories: the first one meets the logarithmic, instead of exponential, relationship, which indicates a distribution of the energy barriers, and the second one is the stepwise magnetization change with time. The latter could be a signature for the nucleation and growth of magnetic phases under applied field similar to martensite domains[50] or the inhomogeneous distribution of substituted rare earth elements in the compounds.[37]

Investigations carried out by Wang et al[40] revealed that the $T_N$ of $La_{1-x}Nd_xFe_{11.5}Al_{1.5}$ decreases from 205 K to 188 K as Nd content increases from x=0 to 0.2. When the Nd content is low, such as x=0.1, lattice distortion is small, and therefore the changes in the Fe-Fe, Fe-Al-Fe bond length and Al-Fe-Al bond angle are small. Meanwhile, the number of the Nd-Fe pairs with an FM coupling is few. As a result, AFM coupling prevails, and the FM clusters appear only in some local regions. With Nd content increasing to x>0.2, Nd-Fe FM pairs grow, and Fe-Fe and Fe-Al-Fe bond length and bond angle change significantly. AFM coupling is destructed and the FM ordering prevails. Liu et al[37] and Wang et al[40] investigated the phase diagram of $La_{0.9}Nd_{0.1}Fe_{11.5}Al_{1.5}$ and found an obvious enhancement of the FM characteristics of the magnetic clusters with applied field. Under a constant field below 3 T, AFM-FM, FM-AFM, AFM-PM transitions appear successively in the warming process. A higher magnetic field causes a lower/higher temperature shift of the AFM-FM/FM-AFM transition, and a slighter decrease of $T_N$. Further investigations indicated that large magnetic entropy change appears around FM-AFM transition. Wang et al[40] investigated magnetic entropy change based on both the magnetic and heat capacity measurements, and found that the results obtained from different techniques are consistent with each other. This fact proved the rationality of using Maxwell relation to calculate entropy change accompanying a first-order phase transition. For a field change of 0-5 T, the maximal ΔS is 9 J/kg K for $La_{0.8}Nd_{0.2}Fe_{11.5}Al_{1.5}$, nearly the same as that of Gd. However, due to the large heat capacity of the $La(Fe_{1-x}Al_x)_{13}$-base compounds, the adiabatic temperature change is only one third of that of Gd. Fortunately, the adverse effect caused by large lattice capacity can be avoided by choosing a suitable mode of refrigeration. For example, refrigerants in the refrigerators based Ericsion cycle are required to have nearly constant magnetocaloric values in a wide temperature range. $La_{0.8}Nd_{0.2}Fe_{11.5}Al_{1.5}$ undergoes two successive transitions at two close temperatures, which results in a nearly constant entropy change in a wide temperature range.

## 5. Effect of introducing interstitial C or H atoms on magnetic and magnetocaloric properties in $LaFe_{13-x}Al_x$

Wang et al[34] and Chen et al[51] investigated the effect of interstitial atom C on the magnetic and magnetocaloric properties of $LaFe_{13-x}Al_xC_y$. It is found that to get stable $NaZn_{13}$-structure the concentration of C generally cannot exceed $y=0.5$ and 0.8 for, respectively, the intermetallics of $x=1.5$ and 1.6, and the introduction of C causes a lattice expansion. Fascinatingly, the introduction of a small amount of C, such as $y=0.1$, can completely convert the AFM ordering in $LaFe_{13-x}Al_x$ into an FM one, and the Curie temperature increases monotonously with the content of C. For example, $T_C$ of $LaFe_{11.5}Al_{1.5}C_y$ increases from 191 K to 262 K as C content changes from $y=0.1$ to $y=0.5$. A linear relation between $T_C$ and lattice parameter is further observed.

Experiments show that the effect of interstitial atoms on $T_C$ is two-fold. First, it leads to reduction of the 3d energy band as a consequence of lattice expansion that reduces the overlap of the Fe3d electron wave functions. As a result, the FM interaction enhances and $T_C$ grows. Second, it will produce a hybridization of the C and Fe electron orbits, thus a decrease of $T_C$. The observed increase of $T_C$ with lattice parameter indicates that the effect of lattice expansion plays a dominant role. Besides, the saturated magnetization slightly increases with increasing C content. The MCE of $LaFe_{11.5}Al_{1.5}C_y$ was studied by magnetic measurements,[34] and the obtained entropy change is comparable to that of Gd, as shown in Fig. 8.

To deeply understand the effect of C introduction on magnetic properties of the AFM $LaFe_{13-x}Al_x$ compounds, Zhang et al[38] prepared $LaFe_{11.4}Al_{1.6}C_y$ (0<$y$≤0.08). Investigations reveal the coexistence of the AFM and FM clusters in the compound of $y$=0.02 and the occurrence of AFM-FM and FM-AFM transitions under external fields. These phenomena are very similar to those observed in $LaFe_{11.4}Al_{1.6-x}Si_x$ ($x$=0.22), $La_{0.8}Pr_{0.2}Fe_{11.4}Al_{1.6}$, and $La_{0.9}Nd_{0.1}Fe_{11.5}Al_{1.5}$. Furthermore, many stepwise magnetization changes were observed around AFM-FM, FM-AFM transitions, manifesting the metastable characteristics of the magnetic state.[52] When $y$=0.04 and 0.06, ground state converts from AFM to FM state, and two FM-AFM and AFM-FM transitions occur successively. With the increase of the content of C, the FM-AFM transition shifts to high temperatures. For the compound of $y$=0.08, the temperature for the FM-AFM transition coincides with the Neel temperature, and the compound displays a simple FM-PM transition.

For the $LaFe_{13-x}Al_x$ with a FM ground state, the introduction of C significantly enhances both the Curie temperature and the magnetic moment of Fe, the latter causes a growth of saturation magnetization. As C content changes from $y$=0 to $y$=0.5, for example, $T_C$ of $LaFe_{11}Al_2C_y$ increases from 200 K to 291 K and Fe magnetic moment from 1.81 $\mu_B$ to 1.93 $\mu_B$. Large MCE was observed around transition temperature for $LaFe_{11}Al_2C_y$, and the peak value of the entropy change is ~6.5 J/kg K under a field change of 0-5 T. Although the peak of the entropy change is not high, its temperature span is wide (~70 K), which gives rise to a quite large refrigeration capacity (RC).[53]

Zhao et al[54] revealed that the introduction of interstitial H atoms can also produce significant effect on magnetic and magnetocaloric properties of $LaFe_{11.5}Al_{1.5}H_y$. As H concentration 0.12≤y≤1.3, The $LaFe_{11.5}Al_{1.5}H_y$ intermetallic with H content between 0.12 and 1.3 displays a FM character, and the Curie temperature increases from ~225 K to ~295 K as H concentration changes from $y$=0.12 to 1.3. The magnetic moment of Fe also shows an increase with the addition of H. It is 1.99 $\mu_B$ for $LaFe_{11.5}Al_{1.5}$, and increases to 2.12 $\mu_B$ and 2.19 $\mu_B$ as $y$ grows to 0.12 and 1.3. We fitted the isothermal magnetization curves to the Inoue-Shimizu model, and obtained the Landau coefficients of $\alpha_2(T_C)$=0.118, −0.063, and −0.373 for $LaFe_{11.5}Al_{1.5}H_x$ ($x$=0.12, 0.6, and 1.3), respectively. This result indicates that the transition nature develops from second-order to first-order with increasing H concentration. Figure 9(a) presents temperature-dependent magnetization for $LaFe_{11.5}Al_{1.5}H_x$ under different magnetic fields. The maximal values of Δ$S$ is 5.4, 6.2, and 6.7 J/kg K under a field change of 0-2 T, and 10.1, 11.6, and 12.3 J/kg K under 0-5 T, for the compounds with $x$=0.12, 0.6, and 1.3, respectively. The entropy change increases by 22% as H concentration increases from 0.12 to 1.3. These results can be ascribed to two effects. One is the strengthening of the first-order nature of the IEM transition as indicated by Landau coefficients $\alpha_2(T)$, which favors enhanced entropy changes. Another is the increase of Fe moment with introducing H atoms, which also enhances the entropy change. However, the interstitial hydrides exhibit a low thermal stability. Further studies found that the simultaneous introduction of C and H atoms can remarkably enhance the thermal stability of $LaFe_{13-x}Al_x$, without affecting the excellent magnetic and magnetocaloric properties. For example, $LaFe_{11.5}Al_{1.5}C_{0.2}H_y$ keeps stable even when heated to 639 K. As H content increases from $y=0$ to 1.0, $T_C$ increases from 212 K to 309 K, and maximal $\Delta S$ increases from 6.7 J/kg K to 8.5 J/kg K, for a field change of 0-2 T, and from 11.9 J/kg K to 13.8 J/kg K, for 0-5 T (Fig. 9(b)). Around room temperature, $LaFe_{11.5}Al_{1.5}C_{0.2}H_{1.0}$ exhibits an entropy change larger than that of Gd by 70% and 42% for a field change of 0-2 T and 0-5 T, respectively. As well known, Gd has been long regarded as an excellent refrigerant for magnetic cooling in the room temperature region.

## 6. Bistable AFM and FM states in $LaFe_{13-x}Al_x$ compounds

Intensive investigations have been carried out for the AFM $LaFe_{13-x}Al_x$ compounds. Wang et al[55] investigated the issues associated with the bistable behaviour of AFM and FM states in $LaFe_{13-x}Al_x$ compounds. They chose $LaFe_{11.4}Al_{1.6}$ and constructed magnetic diagram through measuring temperature/field dependent magnetization *(M-T/M-H* curves) under different magnetic fields/temperatures. $LaFe_{11.4}Al_{1.6}$ shows a Neel temperature at 195 K. Figure 10(a) presents the isothermal magnetization curves of $LaFe_{11.4}Al_{1.6}$, measured at different temperatures. One can notice that at temperatures lower than $T_N$, an AFM-FM first-order phase transition can occur at a critical field of $H_{on}$ for the field increase operation, and a reverse FM-AFM transition takes place at a critical field $H_{off} << H_{on}$ for the field decease running. The hysteresis loop around the metamagnetic transition becomes narrow with the increase of temperature and, finally, disappears at $T_N$. This result evinces the bistable behaviour of AFM and FM states with the change of temperature. Based on the field-dependent magnetization measured at different temperatures, magnetic phase diagram can be constructed in in *H-T* plane (Fig. 10(b)). The *H-T* plane is divided into four regions by the $H_{on}$-*T* and $H_{off}$-*T* curves, corresponding to different magnetic phases. The intermediate section enclosed by the two curves is the AFM region when measured in the field ascending mode and FM region in the field descending mode. The arrows indicate the direction of the applied magnetic field. To get a thorough understanding of the bistable AFM and FM states and the temperature dependence of the critical field, Wang et al also analyzed the thermal magnetization and reciprocating thermal magnetization curves measured under different fields and, based on the data thus obtained, refined the phase diagram of LaFe$_{11.4}$Al$_{1.6}$. It was found that in the case of field increasing bistable AFM and FM states appear in the field region of 2 T<*H*<2.2 T and the temperature range of 75 K<*T*<120. However, the bistable behavior occurs in wider regions in the case of field decreasing, i.e. 1.7 T<*H*<2.2 T, *T*<33 K and 45 K<*T*<100 K. Based on the FM-PM transition temperature obtained from the thermal magnetization curves, the boundary between the FM and PM phases in the *H-T* plane was determined, as denoted by the dashed line in Fig. 10(b). The triple point of LaFe$_{11.4}$Al$_{1.6}$ locates at *T*=150 K, *H*=2.6 T. The bistable AFM and FM states in a wide range are closely related to the crystal structure and magnetic interactions, as well as the change of lattice parameter of LaFe$_{11.4}$Al$_{1.6}$. It is a result of interplay between elastic energy and exchange energy. More specifically, the bistable behaviour is closely related to the stability of the boundary phase that is jointly determined by coupled magnetic interaction and lattice distortions.

**7. Hyperfine interactions during metamagnetic transition in AFM LaFe$_{11.4}$Al$_{1.6}$ compound**

Wang et al[56] measured the $^{57}$Fe Mössbauer spectra of the LaFe$_{11.4}$Al$_{1.6}$ compound under different applied fields, and investigated the change of hyperfine parameters during the metamagnetic transition. Figure 11 presents the $^{57}$Fe Mössbauer spectra measured at 5K and different magnetic fields. The relatively broad peaks indicate a wide distribution of hyperfine fields. Because of the small difference of the hyperfine fields at the Fe$^{I}$ and Fe$^{II}$ sites, it is difficult to separate the total spectra into sub-spectra of Fe$^{I}$ and Fe$^{II}$, and only average hyperfine parameters can be obtained. The specific intensity of the Mössbauer sixline spectra depends on the angle $\beta$ between the incident $\gamma$-ray and effective field. The relatively strong intensity of the second and fifth peaks in the spectra reveals the rotation direction of magnetic moment. $I_{2,5}=0$ indicates that the hyperfine field is parallel to external magnetic field. The 2-5 peaks in the spectra of LaFe$_{11.4}$Al$_{1.6}$, collected at 5 K, gradually disappear with increasing magnetic field, indicating a gradual moment rotation to that of external field. Figure 12(a) presents $I_{2,5}$ intensity as a function of external magnetic field.

Figures 12(b) and 12(c) show, respectively, the average hyperfine field $<H_{hf}>$ and its rotating angle $<\cos\theta>$ as a function of magnetic field. One can find that the critical field for inducing the metamagnetic transition is about 4T for LaFe$_{11.4}$Al$_{1.6}$, consistent with the result from magnetic measurements. If the spatial distribution of Fe moment is a narrow single peak, $\cos\theta$ actually represents the projection of unit magnetic moment along external field. Therefore, Figure 12(c) is actually a description of the average rotation of the magnetic moment of Fe with external magnetic field. Under the impact of external field, the magnetic moment of Fe rotates continuously towards the field direction and, when the applied field exceeds a critical value, the sample exhibits a FM state, and almost all moments are aligned in the field direction. The direction of the magnetic moment varies slightly for further field increase, and a tendency of saturation appear above 8 T. To understand the change of hyperfine field with external field, it is necessary to consider the relationship between hyperfine field and Fe moment. One knows that the hyperfine field can be split up into local and transferred contributions which are directly related to the local moments and to the moments of the neighboring atoms.[57,58] For LaFe$_{11.4}$Al$_{1.6}$, the local moment changes little with magnetic field, thus the change of hyperfine field is mainly caused by the transferred contributions from the neighboring Fe atoms. For the conventional AFM state, due to the symmetric atomic environments around Fe atoms, the exchange polarization contributions from neighboring atoms cancel each other and the transferred contributions disappear. However, in LaFe$_{11.4}$Al$_{1.6}$ the Fe$^{II}$ site has asymmetric local environments since the substitution of Al for Fe has destroyed the originally symmetric neighboring environments of Fe$^{I}$ site. As a result, in the AFM state for H=0 the exchange polarization contributions from neighboring atoms do not completely cancel each other, and the transferred contributions show up. As shown by our experiments, the transferred contribution is about 6.24 T for the AFM ground state of LaFe$_{11.4}$Al$_{1.6}$. It increases with increasing field, and when external field exceeds a critical value, the magnetic moment approach a saturation value. In this case, the contribution from neighboring atoms to exchange polarization reaches maximum and thus the transferred contribution does not change any more. The maximal transferred contribution is estimated to be 10.66 T.

Our experimental results indicate that the hyperfine parameters continuously change with external magnetic field. This result is different from the behavior of first-order transition measured by magnetic measurements. Through calculating field-dependent change of average hyperfine field and its rotating angle, it was found that the critical field for the metamgnetic transition is consistent with that derived from magnetic measurements. The local moment of Fe atom keeps nearly unchanged in the field ascending process, and the local hyperfine field changes little. Thus, the change of hyperfine field is mainly caused by transferred contribution due to spin rotation.

**8. Summary**

(1) With increasing Al content, the $LaFe_{13-x}Al_x$ compounds exhibit AFM ($1.0 \leq x \leq 1.8$), FM ($1.8 < x \leq 4.9$), and mictomagnetic properties ($4.9 \leq x \leq 7.0$) successively. For $LaFe_{13-x}Al_x$ with an AFM ground state, weak AFM properties show up because of the high coordination number around Fe and the short Fe-Fe bond length. An AFM to FM transition is easy to be induced by magnetic field. The AFM-FM transformation can also be produced by the substitution of Si for Al, Co for Fe, and magnetic R (rare earth atoms) for La, or the introduction of interstitial atoms of C and H.

(2) The coexistence of AFM and FM clusters was observed in compounds $LaFe_{11.4}Al_{1.6-x}Si_x(x=0.22)$, $La_{0.8}Pr_{0.2}Fe_{11.4}Al_{1.6}$, $La_{0.9}Nd_{0.1}Fe_{11.5}Al_{1.5}$, and $LaFe_{11.4}Al_{1.6}C_{0.02}$. In a considerably wide temperature range below the Neel temperature, field-induced AFM-FM and FM-AFM transitions take place. Meanwhile, obvious relaxation behavior was observed around the transition temperature. With the increase of magnetic field, the ratio of FM phase grows and the compounds completely transit to the FM state under high magnetic fields.

(3) For compounds $La_{1-x}R_xFe_{11.5}Al_{1.5}$ (R=Ce、Pr、Nd), the content of R generally cannot exceed 30% to keep a stable $NaZn_{13}$-structure. It becomes easier for magnetic field to induce an AFM-FM transition in the presence of R, and the threshold field for the AFM-FM transition decreases considerably with the increase of the R content. The effect of *R* on the magnetic properties of the compounds increases progressively along the sequence of Ce, Pr, and Nd.

(4) Through measuring temperature/field dependent magnetization under different fields/temperatures for LaFe$_{11.4}$Al$_{1.6}$, magnetic diagram was constructed and a metastable state with coexisted AFM and FM phases was observed. Near the critical region in the diagram, the coexistence region is different for field increasing and decreasing. In the case of field increasing the two phases coexist at field region about 2 T<*H*<2.2 T and temperature region about 75 K<*T*<120. But for the case of field decreasing the coexistence occurs at wider regions, i.e. 1.7 T<*H*<2.2 T, *T*<33 K and 45 K<*T*<100 K.

(5) The hyperfine parameters display a continuous change with external magnetic field for LaFe$_{11.4}$Al$_{1.6}$. This phenomenon may be related to the nearly same intrinsic moments in FM and AFM states. Considering the fact that the local moment of Fe atom keeps nearly unchanged upon application of magnetic field and the local hyperfine field changes little, the change of hyperfine field is mainly caused by transferred contribution due to spin rotation.

(6) The Curie temperature of the LaFe$_{13-x}$Al$_x$-based compounds can be promoted to room temperature by introducing Co or interstitial atoms C and H, and large MCE was observed around the transition temperature. The room temperature entropy change of LaFe$_{10.88}$Co$_{0.95}$Al$_{1.17}$ is 4.5, and 9.0 J/kg K for a field change of 0-2 T, and 0-5 T, respectively. Simitanously introducing C and H enhances thermal stability of the compounds. For example, LaFe$_{11.5}$Al$_{1.5}$C$_{0.2}$H$_y$ shows not only an excellent thermal stability around room temperature but also large entropy change. Its entropy change is 8.5 and 13.8 J/kg K for 0-2 T and 0-5 T, respectively, exceeding that of Gd (5.0 and 9.7 J/kg K). It therefore is a potential candidate for magnetic refrigerants.

Figure captions

Fig. 1 Temperature dependence of the magnetic entropy change of La(Fe$_{1-x}$Co$_x$)$_{11.7}$Al$_{1.3}$ ($x$=0.02, 0.04, 0.06 and 0.08) compounds for magnetic field changes of 0−2 T and 0−5 T, respectively.

Fig. 2. (a) Magnetic phase diagram in a magnetic field of 0.1 T for LaFe$_{11.4}$Al$_{1.6-x}$Si$_x$ (0≤$x$≤1.6) compounds. The solid lines and dotted lines indicate magnetic phase boundaries, $T_N$, $T_0$ and $T_C$ indicate antiferromagnetic (AFM)–paramagentic (PM), ferromagnetic (FM)–AFM and FM–PM transition temperatures, respectively and '↕' illustrates the temperature region where FM clusters grow from the AFM background for the compound with $x$=0.22.[44] (b) Temperature dependence of electrical resistivity under different magnetic fields and the corresponding magnetization in a magnetic field of 0.1 T for LaFe$_{11.4}$Al$_{1.6-x}$Si$_x$ ($x$=0.22) compounds.

Fig. 3 Temperature dependence of the magnetic entropy change of LaFe$_{11.4}$Al$_{1.6-x}$Si$_x$ ($x$=0.3 (a), 0.8 (b)) compounds for different magnetic field changes up to $H$ = 5 T.[44]

Fig. 4 Temperature dependence of the magnetic entropy change of LaFe$_{11.03}$Co$_{0.8}$Al$_{1.17-x}$Si$_x$ ($x$=0.47, 0.70, 0.94 and 1.17) compounds for different magnetic field changes up to $H$ = 5 T.[47]

Fig. 5 Isothermal magnetization curves of La$_{1-x}$R$_x$Fe$_{11.5}$Al$_{1.5}$ ($R$=Ce, Pr and Nd) compounds at 5 K in the field-ascending and field-descending processes.[48]

Fig. 6 (a) Temperature dependence of the magnetization for $La_{0.8}Pr_{0.2}Fe_{11.4}Al_{1.6}$ measured in ZFC, FCC and FCW modes under a magnetic field of 0.45 T.[49] (b) Temperature dependence of the magnetization for $La_{0.9}Nd_{0.1}Fe_{11.5}Al_{1.5}$ compound under different magnetic fields.[37]

Fig. 7 Reduced magnetization ($M/M_0$) versus time ($t$) for $La_{0.8}Pr_{0.2}Fe_{11.4}Al_{1.6}$ measured at typical temperatures under $H = 0.45$ T in the ZFC mode.[49] At each $T$, $M_0$ is the value of the magnetization recorded when the relaxation measurements were started, i.e. 0 s after the target $H$ and $T$ values were reached.

Fig. 8 Temperature dependence of the magnetic entropy change of $LaFe_{11.5}Al_{1.5}C_y$ ($y$=0.1, 0.2, 0.4 and 0.5) compounds compared to that of Gd for magnetic field changes of 0–2 T and 0–5 T, respectively.[34]

Fig. 9 Temperature dependence of the magnetic entropy change under magnetic field changes of 0–2 T and 0–5 T, for (a) $LaFe_{11.5}Al_{1.5}H_x$ ($x$=0.12, 0.6 and 1.3),[54] and (b) $LaFe_{11.5}Al_{1.5}C_{0.2}H_x$ ($x$=0, 0.5 and 1.0), respectively.

Fig. 10 (a) Isothermal magnetization curves of $LaFe_{11.4}Al_{1.6}$ compound at different temperatures in the field-ascending and field-descending processes. (b) Magnetic phase diagram in $H$-$T$ for $LaFe_{11.4}Al_{1.6}$ compound.[55]

Fig. 11 $^{57}$Fe Mössbauer spectra at 5 K in various applied fields for $LaFe_{11.4}Al_{1.6}$.[56] The best fit is shown by the solid line.

Fig. 12 Applied field dependences of (a) the relative intensity of the second and fifth lines $I_{2,5}$, (b) the average hyperfine field $H_{hf}$, and (c) the calculated average $<\cos\theta>$ for $LaFe_{11.4}Al_{1.6}$.[56]

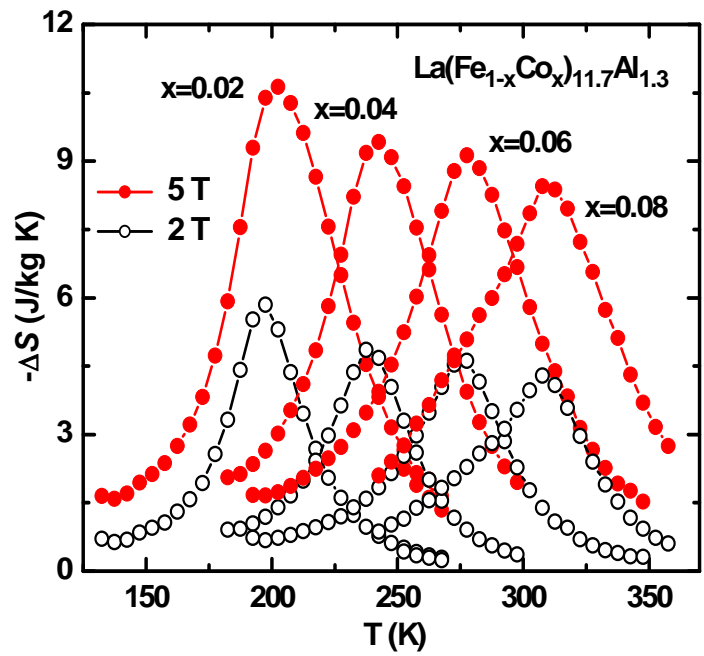

SBG Fig. 1

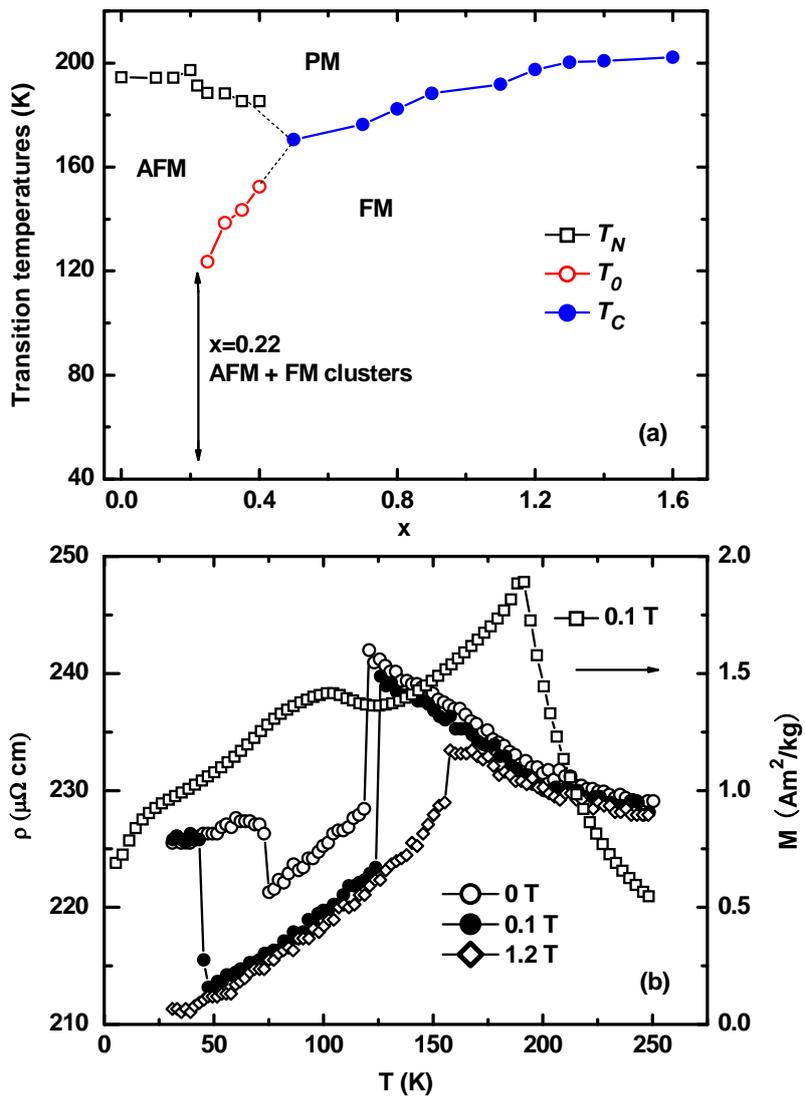

SBG Fig. 2

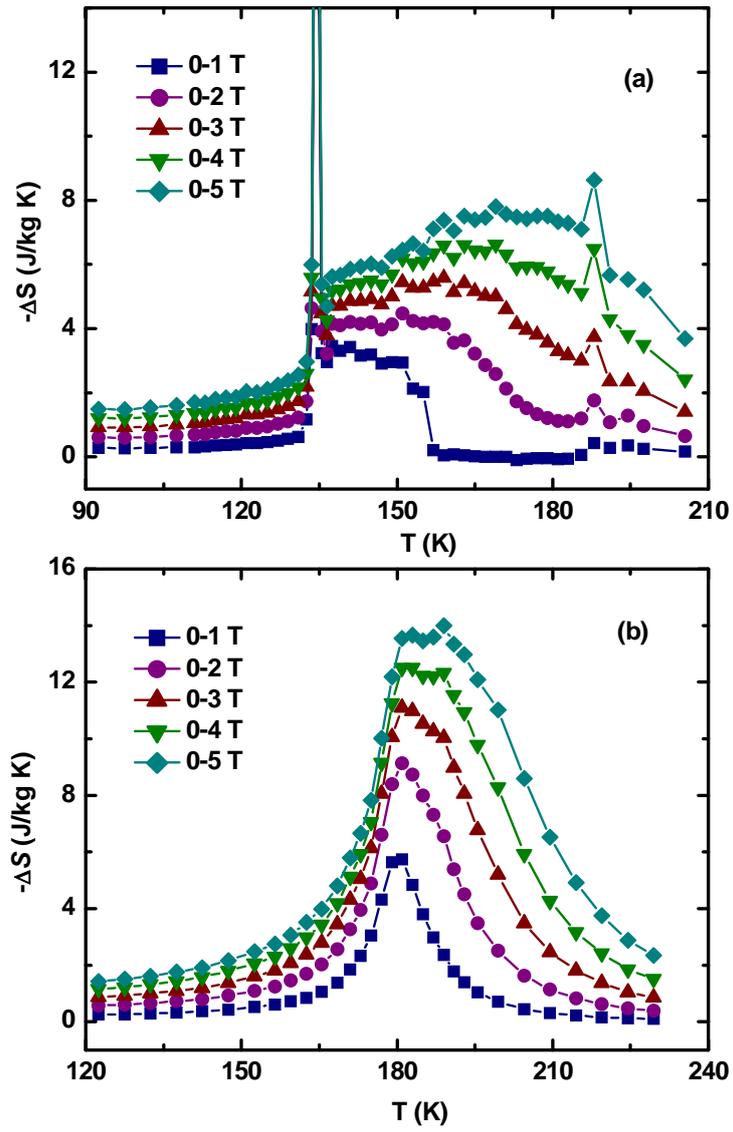

SBG Fig. 3

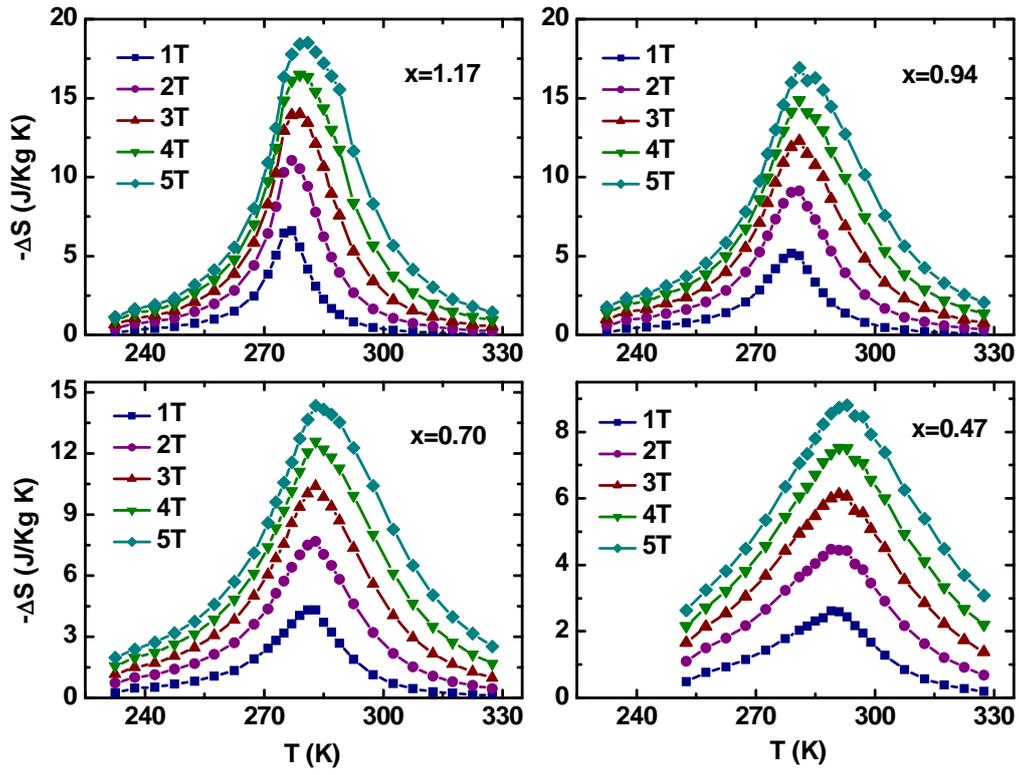

SBG Fig. 4

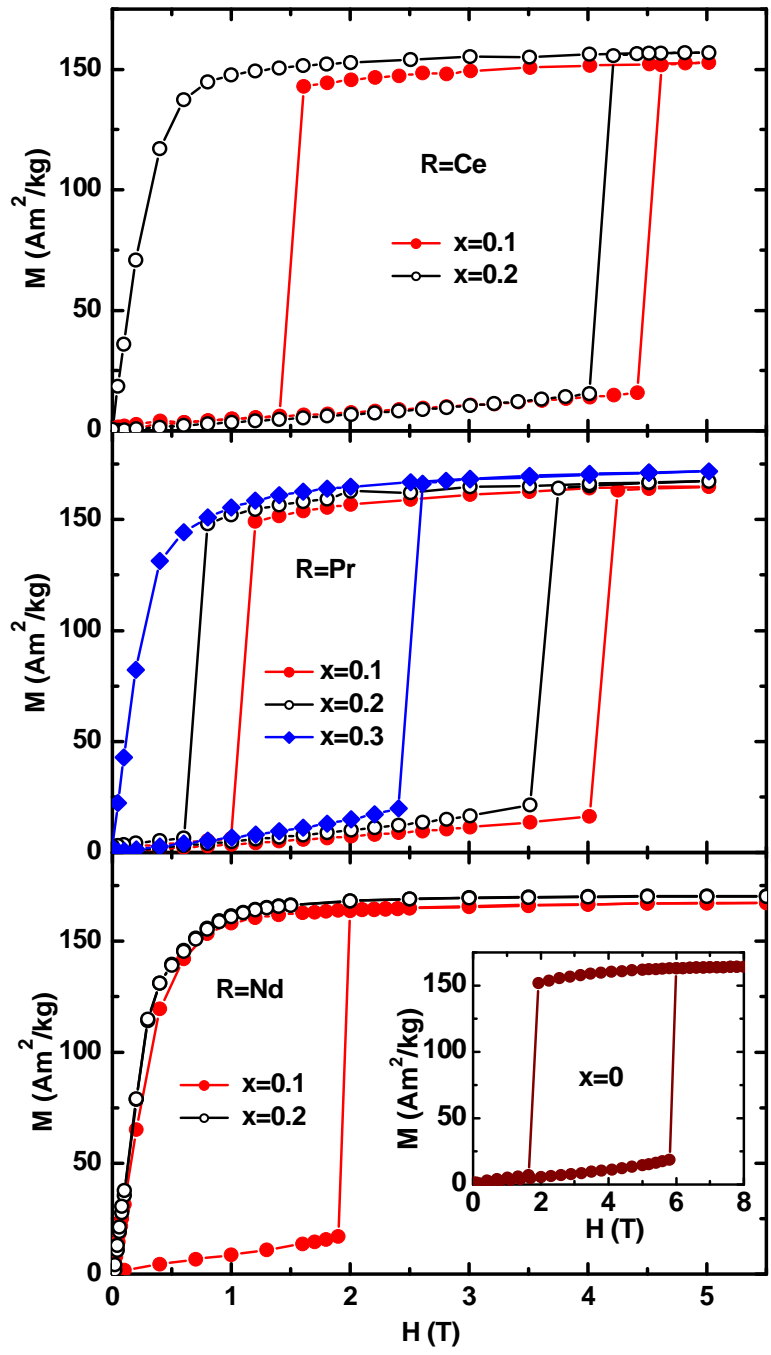

SBG Fig. 5

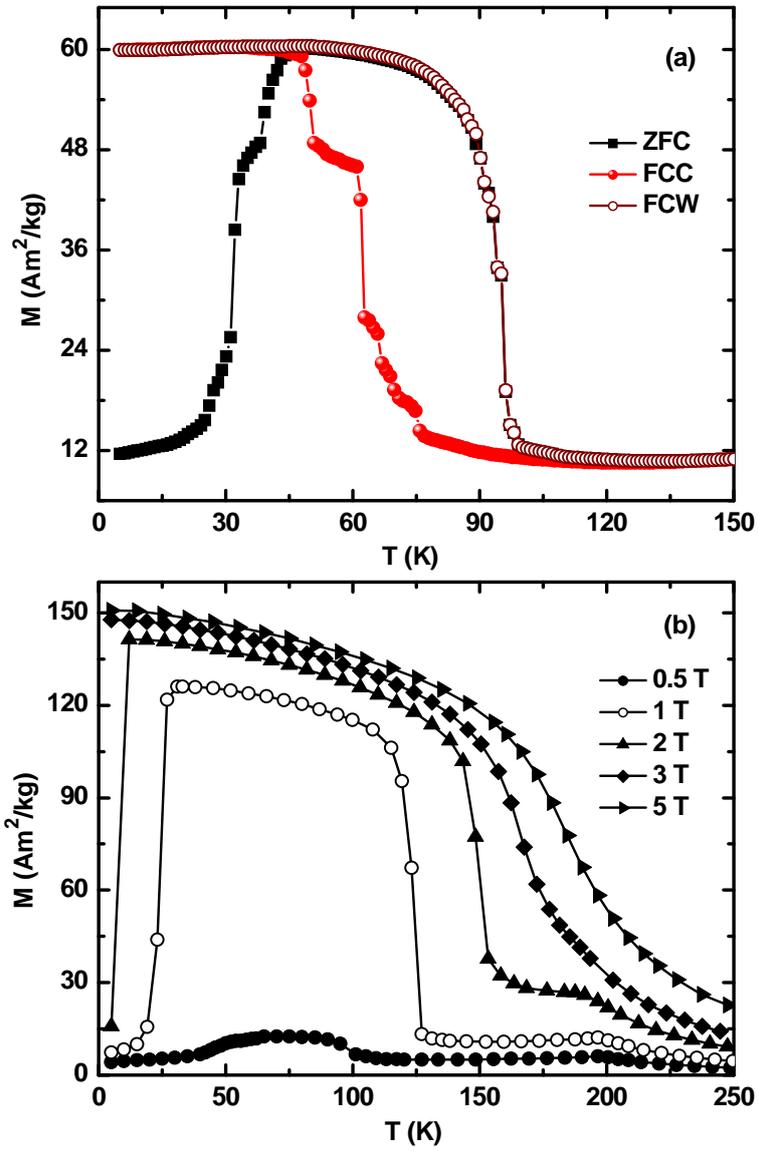

SBG Fig. 6

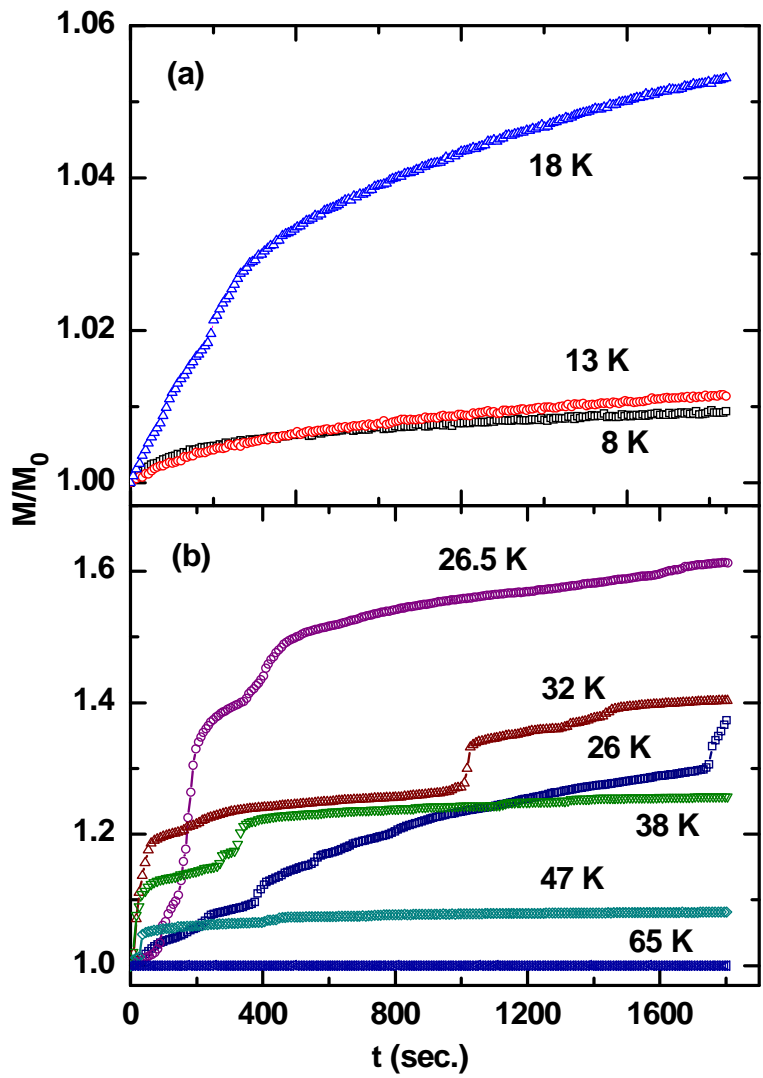

SBG Fig. 7

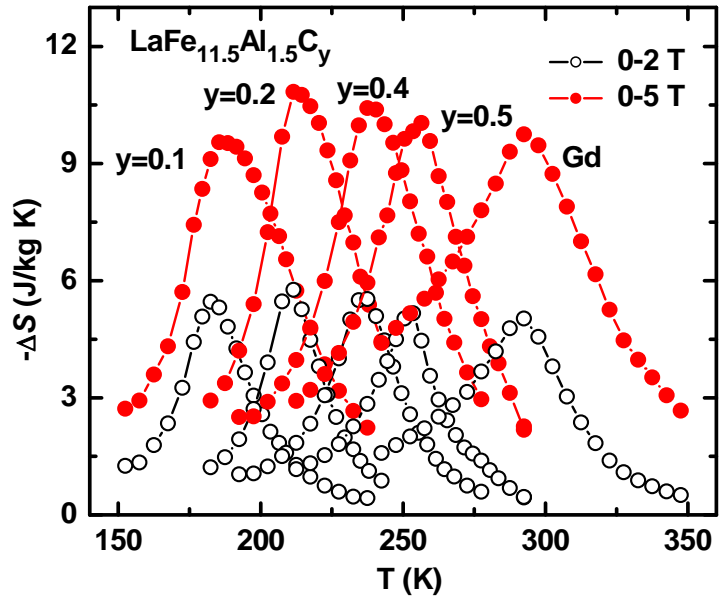

SBG Fig. 8

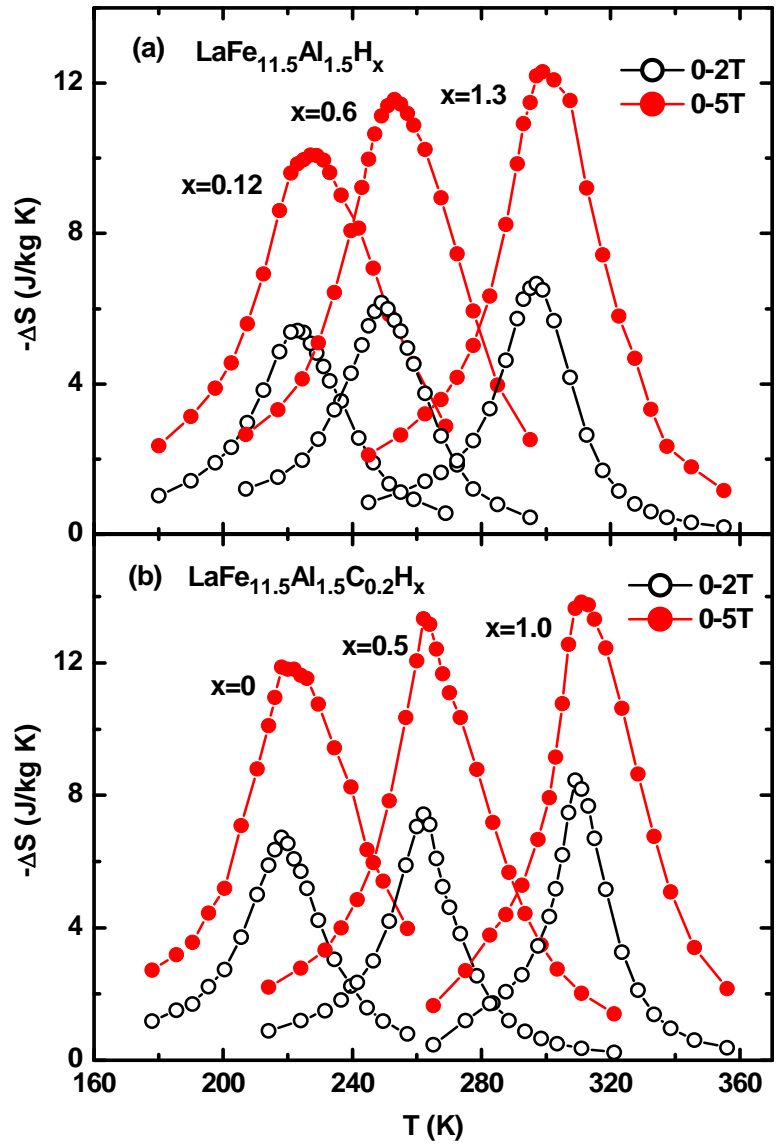

SBG Fig. 9

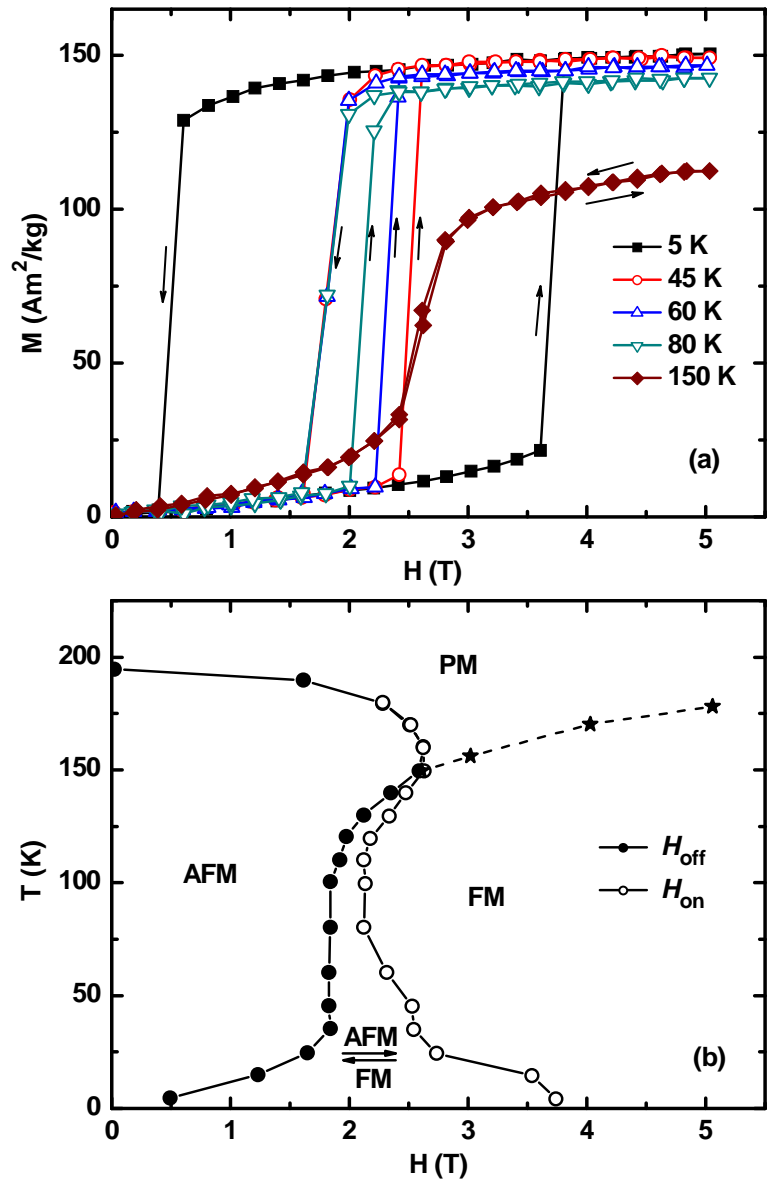

SBG Fig. 10

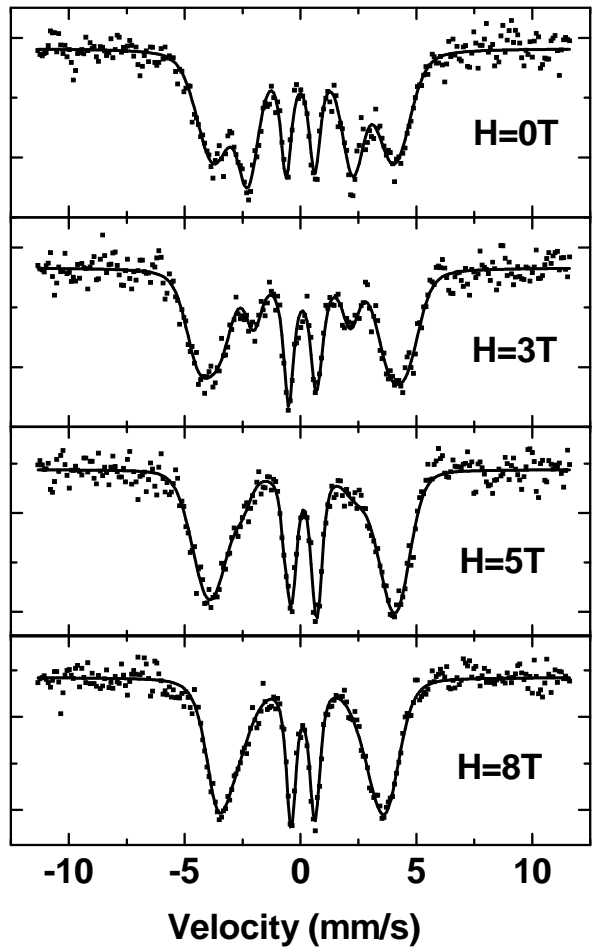

SBG Fig. 11

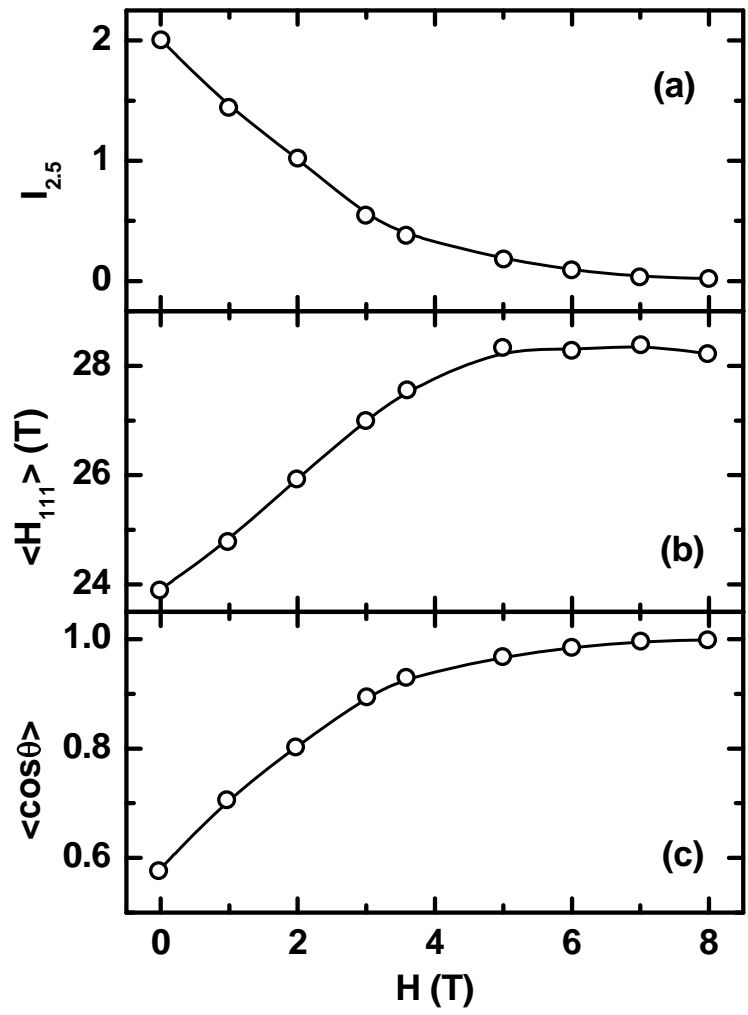

SBG Fig. 12